\documentclass[fleqn,10pt,twocolumn]{wlscirep}
\usepackage{graphicx}
\usepackage{amsmath,amssymb,bm} 
\usepackage[utf8]{inputenc}
\usepackage{fontenc}
\usepackage{lineno}
\usepackage[thinlines]{easytable}
\usepackage{makecell}

\usepackage{fixmath}

\usepackage[final]{changes}
\newcommand{\stkout}[1]{\ifmmode\text{\sout{\ensuremath{#1}}}\else\sout{#1}\fi}
\setdeletedmarkup{\stkout{#1}}

\newcommand{\mbd}{\mathbold}
\newcommand{\nn}{\nonumber \\}

\title{The scaled-invariant Planckian metal and quantum criticality in Ce$_{1-x}$Nd$_x$CoIn$_5$}

\author[1,2]{Yung-Yeh Chang}
\author[3,4]{Hechang Lei}
\author[3,$\dag$]{C. Petrovic}
\author[1,2,$\ddag$]{Chung-Hou Chung}

\affil[1]{Physics Division, National
Center for Theoretical Sciences, Taipei 10617, Taiwan, Republic of China}
\affil[2]{Department of Electrophysics, National Yang-Ming  Chiao-Tung University, Hsinchu, 300 Taiwan, R.O.C.}
\affil[3]{Condensed Matter Physics and Materials Science Department, Brookhaven National Laboratory, Upton, New York 11973-5000, USA}
\affil[4]{Present Address: Department of Physics and Beijing Key Laboratory of Opto-electronic Functional Materials \& Micro-nano Devices, Renmin University of China, Beijing, China}

\affil[$\dag$]{Corresponding author: petrovic@bnl.gov}
\affil[$\ddag$]{Corresponding author: chung0523@nycu.edu.tw}

\begin{abstract}
{Perfect $T$-linear resistivity associated with universal scattering rate: $1/\tau =\alpha  k_B T/(h/2\pi)$ with $\alpha \sim 1$, so-called Planckian metal state,
has been observed in the normal state of a variety of strongly correlated superconductors close to a quantum critical point.
However, the microscopic origin of this intriguing phenomena and its link to quantum criticality still remains an outstanding open problem. In this work, we observe the quantum-critical $T/B$-scaling of the Planckian metal state in the resistivity and heat capacity of heavy-electron superconductor Ce$_{1-x}$Nd$_{x}$CoIn$_5$ in magnetic fields near the edge of antiferromagnetism, \added{driven by critical Kondo hybridization} at the critical doping $x_c \sim 0.03$. \deleted{Clear experimental evidences are shown to support the notion of Kondo hybridization being quantum critical at $x_c$ from (i) the perfect linear relations among Kondo coherence scale,  inverse of slope in $T$-linear resistivity and $1/\alpha$ as a function of doping, (ii) various quantum critical scalings in Planckian state observed near $x_c$.} We further provide the first microscopic mechanism to account for the Planckian state in a quantum critical system based on the critical charge fluctuations near Kondo breakdown transition at $x_c$ within the quasi-two-dimensional Kondo-Heisenberg lattice model. This mechanism simultaneously captures the observed \added{universal}  Planckian \deleted{limit of the universal $T$-} scattering rate as well as the quantum-critical scaling and power-law divergence in thermodynamic observables near criticality. Our mechanism is generic to Planckian metal states in a variety of quantum critical superconductors near Kondo destruction.
}  
\end{abstract}

\begin{document}

\flushbottom
\maketitle
%
%
\thispagestyle{empty}

Metallic behavior that goes against the Landau's Fermi liquid paradigm for ordinary metals has commonly been observed in a wide variety of strongly interacting quantum materials and yet the emergence of such metals is poorly understood. This unconventional metallic or non-Fermi liquid (NFL) behavior often exists near a magnetic quantum phase transition, and shows ``strange metal (SM)" phenomena with (quasi-)linear-in-temperature resistivity and singular logarithmic-in-temperature specific heat coefficient \cite{Lohneysen-RMP, 2019-Sun-CePdAl}. 

The Planckian metal state constitutes a particularly intriguing class of SM  states and has been observed in the normal state of unconventional superconductors, including cuprate superconductors \cite{Taillefer-planckian-2019}, iron pnictides and chalcogenides \cite{Hussey-pnictide0-RPP, Hussey-pnictide-Nature,Kasahara-PRB-pnictide,Jiang-JPCM-pnictide, FeSeS-PRR-2020}, organic \cite{Taillefer-Organic-PRB, Mackenzie-Science} and heavy-fermion compounds \cite{Gegenwart-YRS-PRL,Custers2003Nature, Custers2010prl, Mackenzie-Science}, and twisted bilayer graphene \cite{Cao-SM-TBG-PRL}. The heart of this puzzling state  is that the universal $T$-linear scattering rate reaches the Planckian dissipation limit allowed by quantum mechanics, $1/\tau = \alpha k_B T/(h/2\pi)$ with $\alpha \sim 1$ \cite{Zaanen-2004-nature,Mackenzie-Science,Hartnoll-2015-nature,hartnoll-planckian}. This intriguing observation leads to fundamental questions: Is $\alpha$ a universal constant independent of microscopic details \cite{Patel-SM-PRL}? It was estimated that the Planckian bound can be reached in the low-temperature strong-coupling limit of electron-phonon and marginal Fermi liquid systems \cite{hartnoll-planckian}.  In quantum critical systems, however, it is not clear if and how $\alpha$ depends on microscopic coupling constant at quantum critical point (QCP) and its link to the quantum critical scaling in observables. It was argued that in quantum critical systems with $T$-linear scattering rate, the Planckian time $\tau_{Pl} \equiv (h/2\pi)/k_B T$ is reached naturally ($\alpha \sim 1$) as a result of the scaled-invariant observables $F(h \omega /2\pi k_B T) = F(\omega  \tau_{Pl})$ at criticality \cite{hartnoll-planckian}. However, such expectation has not yet been confirmed by a microscopic mechanism in any quantum critical systems.

\begin{figure*}[ht]
    \centering
    \includegraphics[width=0.95\textwidth]{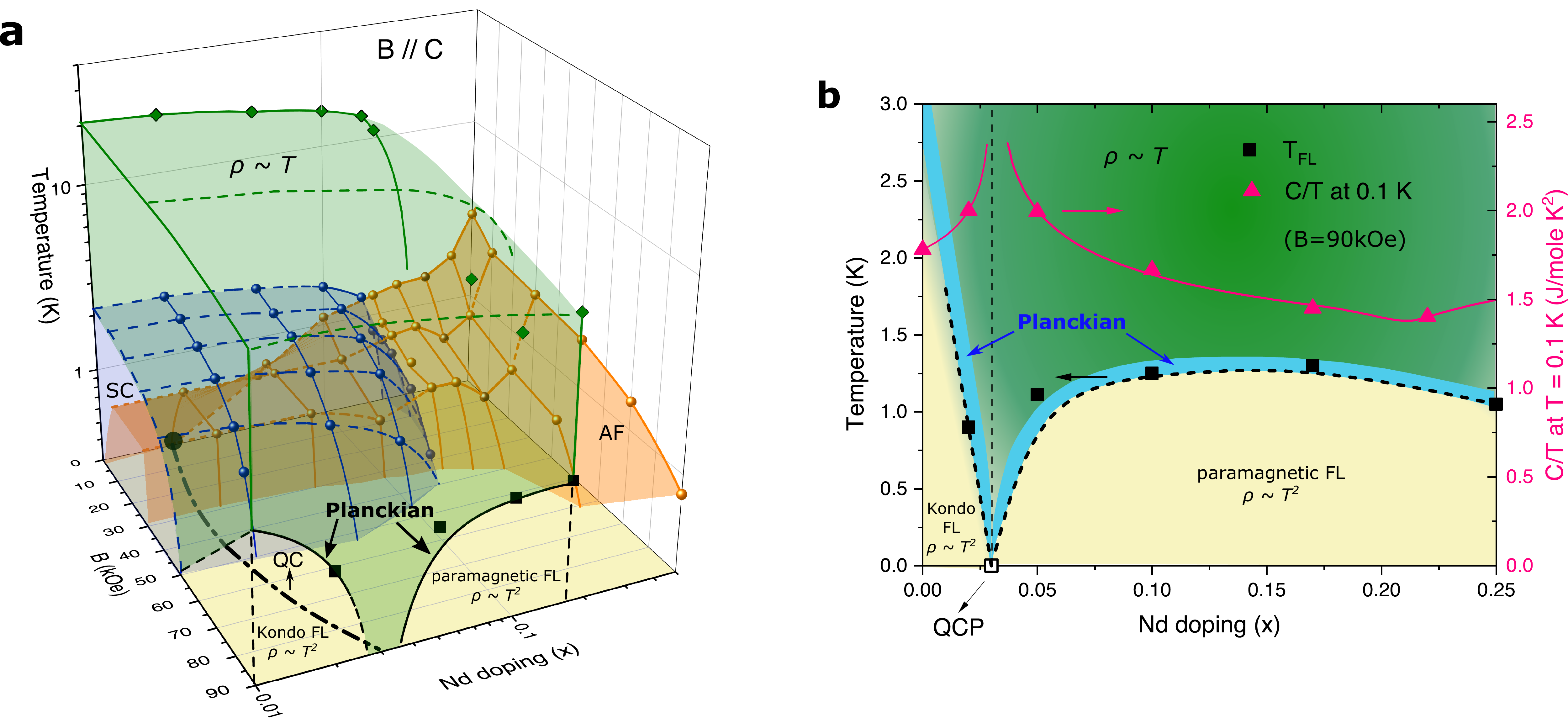}
    \caption{\textbf{The phase diagram for Ce$_{1-x}$Nd$_x$CoIn$_5$ for $B \parallel c$.}  \textbf{a} The ($x,\,B,\,T$) phase diagram: the antiferromagnetic (AF) N\'eel state occurs below the orange area and the superconducting (SC) phase sets in under the blue area (symbols). The data of zero-field plane is reproduced from Fig. 6 of Ref. \cite{Petrovic-PBR-NdCeCoIn}. A QCP is predicted on the doping axis near $x \approx 0.03$ at zero field plane, hidden beneath the coexisting phase of the AF and SC phases. At temperature above the SC and AF phases and below the green region, this material shows linear-in-$T$ strange-metallic (SM) behavior. A quantum critical (QC) line (black dotted-dashed line) is expected to exist at finite field and Nd doping, connected to the QCP at zero field, $x=x_c$.  \textbf{b} Phase diagram of the system at $B=90$kOe. As the external field is large enough (approximately $B \sim 70$ kOe for $B\parallel c$), a QCP emerges at $x\approx 0.03$ as inferred from the $T^2$ resistivity measurement (below $T_{FL}$). This QCP is located at $B=90$kOe along the QC line of \textbf{a}. The enhancement of the normal-state electronic specific heat $C_V/T$ at the low-temperature limit ($T = 0.1$K) is fitted by a power-law singular function  at $x =x_c \approx 0.03$,  $C_V/T|_{T=0.1K} \sim |x-x_c|^{-\beta}$, with $\beta \approx 0.11$ for $x<x_c$ and $\beta \approx 0.16$ for $x>x_c$ (pink curves, right axis). Data of \textbf{b} is measured at $B = 90$kOe. The Planckian scattering rate is reached  in the vicinity of $T_{FL}$ (blue area).}
    \label{fig:PD}
\end{figure*}

While it is challenging to address these questions in cuprates, pnictide and organic superconductors, significant experimental and theoretical progress have  been made in heavy-fermion quantum critical superconductors \cite{Coleman-lecture}. A prototypical example of such systems is the Ce$M$In$_5$ family with $M$ = Co, Rh, Ir under field and pressure where superconductivity and non-Fermi liquid (NFL) normal state properties emerge due to competition between antiferromagnetism and Kondo correlation near a antiferromagnetic Kondo-breakdown (AF-KB) QCP \cite{Petrovic-JPCM,Hegger-CeRhIn5,Petrovic-europhys, Petrovic-PRL-cecoin5, Thompson-SC-cecoin5,Thompson-PRL-2011}. In particular, the $T$-linear resistivity in CeCoIn$_5$ was reported to exhibit Planckian dissipation ($ \alpha \sim 1$) \cite{Mackenzie-Science},  and therefore the Co-115 family is well suited for this study. In this work, we focus on Nd-doped CeCoIn5\cite{Petrovic-PBR-NdCeCoIn}. 
\begin{figure}[ht]
    \centering
    \includegraphics[width=0.4\textwidth]{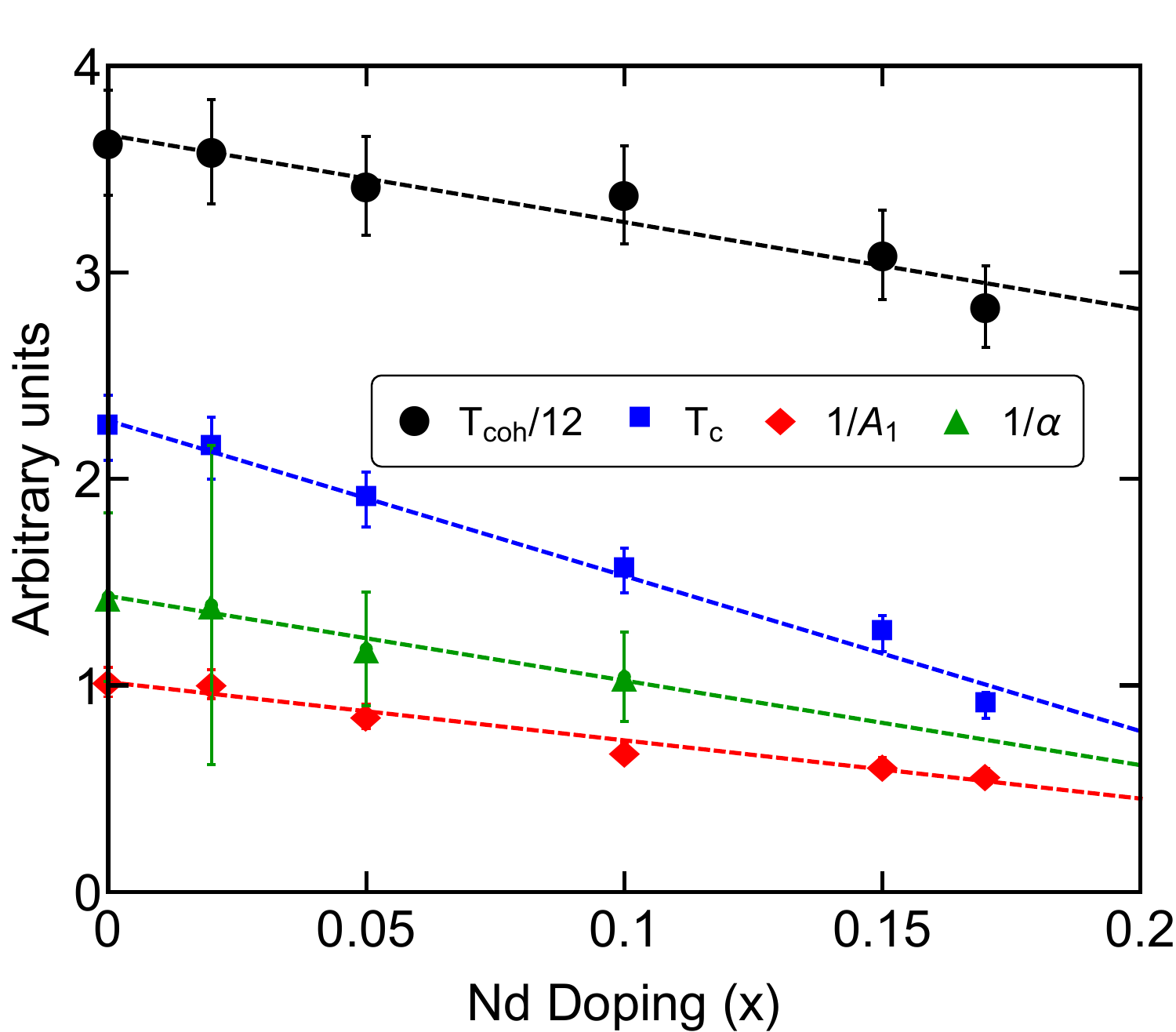}
    \caption{Linear-in-$x$ dependence of $T_{\text{coh}}$, $T_c$ and $1/A_1$, and $1/\alpha$. The data are taken from Ref.\cite{Petrovic-PBR-NdCeCoIn} and Table \ref{eq:table-alpha}.}
    \label{fig:parameter}
\end{figure}

\begin{table*}
\def\arraystretch{1.6}
\centering
\caption{\textbf{The band-specific parameters of Nd$_x$Ce$_{1-x}$CoIn$_5$}. The averaged dHvA frequencies $F$, carrier concentration $n$, and effective masses $m^\star$ of the $\alpha$ band shown in the table are taken from Ref. \cite{Petrovic-PRB-FS}. Carrier concentration $n$ of the $\alpha$-band for $x = 0.05$ is estimated by a $d$-dimensional Fermi volume  with $d = 2.45$, while it is estimated  by a two-dimensional Fermi volume for $x=0, \, 0.02$, in accordance with the angular dependence of the dHvA frequencies measured in Ref.\cite{Petrovic-PBR-NdCeCoIn} . Here, $m_0$ denotes the bare electron mass, while $\alpha$ and $A_{1}$ are extracted from the linear-in-temperature resistivity at zero field from Ref. \cite{Petrovic-PBR-NdCeCoIn}. Note that the $\alpha$ coefficient shown here contains a sub-leading contribution from the $\beta$ band, estimated by the band parameters for pure CeCoIn$_5$ in Ref.\cite{Mackenzie-Science}: $n \approx 0.63\times 10^{28}$m$^{-3}$ (corresponding to the average dHvA frequency $F \approx 9.75$ kT) and $m^\star \approx 100m_0$ (See Supplementary Notes, section \ref{supp:esti-alpha} for details).}
\begin{tabular}{l c c c c c c c}
\hline
&  &  \thead{$x = 0^{\circ} $ \\ ($\theta=0$)}  & \thead{$x = 0.02$ \\ ($\theta=3^{\circ} $)} & \thead{$x = 0.05$ \\ ($\theta=2^{\circ} $)}  & \thead{$x = 0.1$ \\ ($\theta=7^{\circ} $)}  \tabularnewline
\hline 
\hline
$F(kT)~(\alpha\, \text{band})$ &  &  4.9 $\pm$ 0.5   & 4.89 $\pm$ 1.02 & 4.88 $\pm$ 0.5   & 4.41  $\pm$ 0.41 \tabularnewline
\hline 
$n (\times10^{28}m^{-3})~(\alpha\, \text{band})$ &  &  0.31 $\pm$ 0.04  & 0.32 $\pm$ 0.07  & 0.26 $\pm$ 0.02 & 0.17 $\pm$ 0.02  \tabularnewline
\hline 
$m^{\star}(m_{0})~(\alpha\, \text{band})$ &   &  11.7  $\pm$ 2.6  &   11.7 $\pm$ 4.3 &   9.15 $\pm$ 2.0 &   7 $\pm$ 1.0  
\tabularnewline
\hline
$A_{1}(\mu\Omega\cdot cm/K)$   & &  0.98 $\pm$ 0.07 &   1.0 $\pm$ 0.07 &   1.17 $\pm$ 0.08 &   1.49$\pm 0.1$  \tabularnewline
\hline \hline
$\alpha~(\alpha \, \text{band}+ \beta \, \text{band}) $ &  &  0.7 $\pm 0.2$ & 0.72 $\pm 0.4$  & 0.85 $\pm$ 0.2   & 0.96$\pm$ 0.2  \tabularnewline
\hline 
\end{tabular}
\label{eq:table-alpha}
\end{table*}
Single crystals Ce$_{1-x}$Nd$_x$CoIn$_5$  were grown and characterized in Ref. \cite{Petrovic-PBR-NdCeCoIn}. Previous studies showed that antiferromagnetism coexists with superconductivity for $0.02<x<0.17$ \cite{Petrovic-PBR-NdCeCoIn}. Here, we investigate further the Planckian state by applying a magnetic field in this material close to the edge of antiferromagnetism at $x \approx 0.02$. The electrical resistivity and heat capacity measurements were performed in a quantum design PPMS-9 system.  The doping-field-temperature $(x,B,T)$ phase diagram of this material is shown in Fig. \ref{fig:PD}a. Doping Nd into the pure CeCoIn$_5$ effectively introduces chemical pressure and reduces Kondo coupling, therefore favors the long-range antiferromagnetism \cite{Petrovic-PBR-NdCeCoIn, Thompson-PRL-2011}. At zero magnetic field, pure $d$-wave superconducting ground state exists for very low Nd doping $0<x\leq0.02$ \cite{davis-QPI-115}, followed by a co-existing antiferromagnetic superconducting state for $0.05 \leq x< 0.17$, and the long-ranged antiferromagnetic phase is reached for $x>0.17$ \cite{Petrovic-PBR-NdCeCoIn}. For $0<x<0.17$, the resistivity at finite temperatures shows a maximum at temperature $T_{\text{coh}} \sim 45$K where coherent Kondo hybridization is reached, a generic feature of many heavy-fermion superconductors, it then drops to zero at $1$K$<T_c<2$K where superconductivity emerges \cite{Petrovic-JPCM,Petrovic-PBR-NdCeCoIn}. 

The SM behavior with linear-in-$T$ resistivity was observed in the intermediate temperature range $T_c<T<20$K: $\rho(T) = \rho_0+A_1 T$ with $\rho_0$ being residual resistivity extrapolated to zero temperature and $A_1(\alpha)$ being the slope [Fig. S\ref{fig:fig1}a]. Via quantum oscillation experiments in Ref. \cite{Mackenzie-Science}, the scattering rate $1/\tau$ for $x=0$ extracted via Drude formula, $\rho= m^\star/n e^2 \tau $, from the resistivity data combined with the carrier concentration $n_i$, electron effective mass $m^\star_i$ associated with the band $i=\alpha,\, \beta$-$band$ in the $T$-linear resistivity region has been shown to reach the Planckian dissipation limit, $\alpha = (h/2\pi) /k_B  T \tau = (e^2 \rho/k_B T) \sum_i n_i /m^\star_i = A_1 (e^2/k_B) \sum_i (n_i/m^\star_i) \sim 1$. In Table \ref{eq:table-alpha}, we find  similar values of $\alpha \sim O(1)$ for $x=0, 0.02, 0.05$ by  a distinct set of quantum oscillation measurements in Ref. \cite{Petrovic-PRB-FS} (see detailed analysis in Supplementary Notes, section \ref{supp:esti-alpha}). Interestingly, as shown in Fig. \ref{fig:parameter}, we find the inverse slope $1/A_1$, $T_c$ and $T_\text{coh}$ all show linear dependence on $x$ over $0<x<0.15$. In particular, $T_{\text{coh}}$, depending mostly on the Kondo scale, is well approximated by a perfect linear relation to the inverse slope  $1/A_1$ as well as $\alpha,\, T_\text{coh} \propto 1/A_1 \propto 1/\alpha$. Moreover, clear quantum-critical temperature-to-field-scalings in resistivity and specific heat coefficient in Planckian state are observed near $x_c$ (see below).
 These observations strongly suggest (i) Kondo hybridization is quantum critical at $x_c$, and (ii) $\alpha $ depends on the strength of Kondo hybridization in the Planckian SM region, as the second possible scenario mentioned above. To further investigate how $\alpha$ depends on Kondo hybridization near the critical doping $x_c $ and its relation to the possible QCP hidden inside the superconducting dome, we apply an external magnetic field and study signatures of quantum critical scaling in resistivity and specific heat coefficient.

\setlength{\arrayrulewidth}{0.2mm}
\setlength{\tabcolsep}{15pt}


\begin{figure*}[t]
    \centering
    \includegraphics[width=0.95\textwidth]{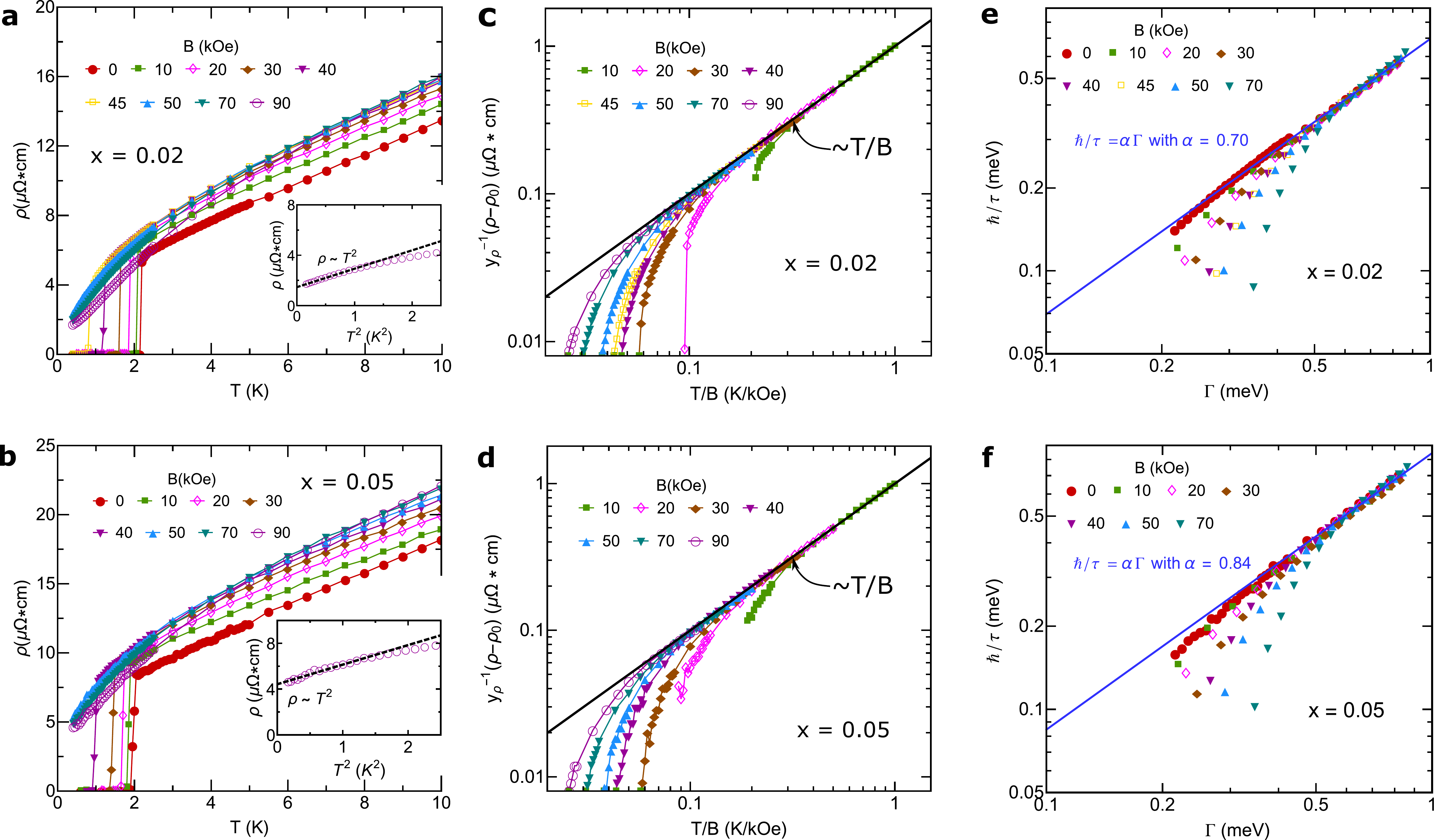}
    \caption{\textbf{Electrical resistivity and the scaling}. \textbf{a. b.} The resistivity $\rho(T)$ of Ce$_{1-x}$Nd$_x$CoIn$_5$ with $x=0.02$ and $x = 0.05$, respectively, for $B \parallel c$. The insets of \textbf{a, b} show the Fermi-liquid behavior with $\rho(T)\sim T^2$ electrical resistivity at low temperatures for $B = 90$ kOe. \textbf{c. d.} The $T/B$ scaling of both \textbf{a} and \textbf{b} show an universal function linearly proportional to $T/B$ (black solid lines).  \textbf{e. f.} The  scattering rates $(h/2\pi)/\tau$ for $x = 0.02,\,\,0.05$ at the intermediate temperature regime where the electrical resistivity exhibits a  linear-in-$T$ dependence  show an universal scaling function $(h/2\pi)/\tau = \alpha \Gamma$ (blue solid lines), with $\Gamma \equiv \sqrt{(k_BT)^2 + (l \mu_B B)^2}$, $\tau$ being the relaxation time, $\mu_B$ being the Bohr magneton, and $l = 0.67$ being a fitting parameter  \cite{PhysComm-pnictide-2020}. In \textbf{e} and \textbf{f}, the $\beta$ band of pure CeCoIn$_5$, with the same band parameters used in Table \ref{eq:table-alpha}, is taken into account in performing the $\Gamma$ scaling of resistivity. The extracted $\alpha$ coefficients shown in \textbf{e} and \textbf{f} are consistent with that shown in Table \ref{eq:table-alpha}.}
    \label{fig:rho-scaling}
\end{figure*}
\begin{figure*}[t]
    \centering 
    \includegraphics[width=0.75\textwidth]{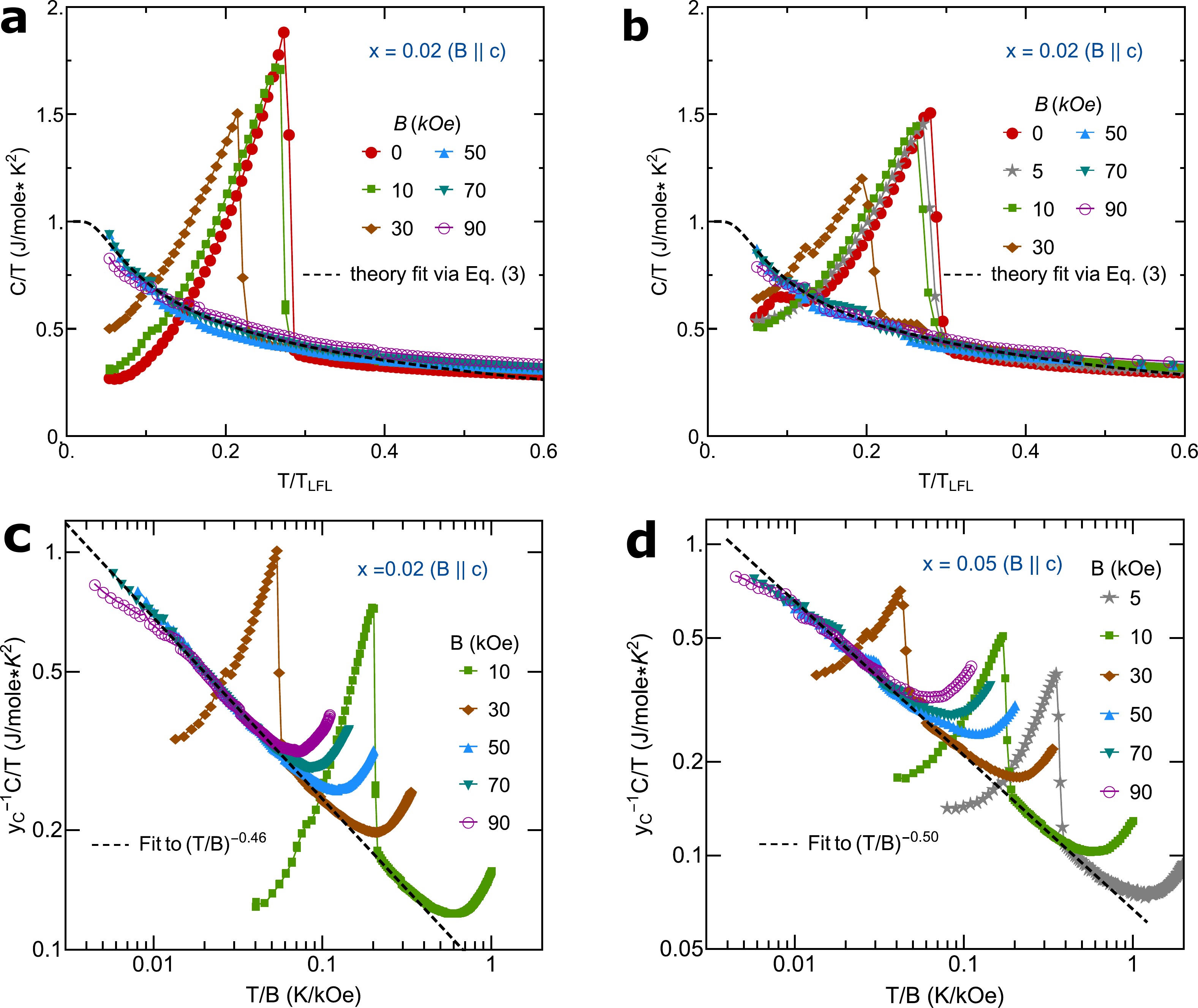}
    \caption{\textbf{Specific heat coefficient $C_V/T$ and its $T/B$ scaling.} \textbf{a} and \textbf{b} show the electronic specific heat coefficient $C_V/T$ versus $T/T_{LFL}$  with different fields $B \parallel c$ for $0.02,$ and $0.05$, respectively while \textbf{c} and \textbf{d} display the power-law $T/B$ scaling of \textbf{a} and \textbf{b}. The dashed lines in \textbf{a} and \textbf{b} are the theory fits to the data via Eq. (\ref{eq:gamma-critical}). Here, $y_C$ in \textbf{c} and \textbf{d} represents a non-universal scaling parameter.}
    \label{fig:Gamma-scaling}
\end{figure*}

As shown in Figs. \ref{fig:rho-scaling}a and \ref{fig:rho-scaling}b, for $x = 0.02$ and $x = 0.05$, the linear-in-$T$ resistivity shows almost the same slopes $(A_1)$ for $0\leq B \leq 70\text{kOe}$. This strongly indicates that the Kondo hybridization strength approximately remains at a constant quantum critical  ``fixed point'' value, associated with a QCP for $0.02< x_c<0.05$ at zero field or a quantum critical line for the above field range (see Fig. \ref{fig:PD}a). The scenario  of the QCP associated with the critical Kondo hybridization is further supported by the following evidences: (i) quantum-critical $T/B$ scaling behavior in the linear-in-$T$ resistivity regime [Figs. \ref{fig:rho-scaling}c and \ref{fig:rho-scaling}d], (ii) the universal Planckian scaling of electron scattering rate $(h/2\pi )/\tau = \alpha \Gamma$ with $\Gamma = \sqrt{(k_B T)^2+(l \mu_B B)^2}$ with $\alpha \sim O(1)$ (see Figs. \ref{fig:rho-scaling}e and \ref{fig:rho-scaling}f), (iii) the $T/T_{LFL}$-scaling in specific heat coefficient $\gamma(T/T_{LFL}) \equiv C_V/T$ (Figs. \ref{fig:Gamma-scaling}\textbf{a} and \ref{fig:Gamma-scaling}\textbf{b}) with $T_{LFL}$ being the Fermi-liquid crossover scale, and (iv) the $T/B$-power-law scaling in the normal state $\gamma$-coefficient: $\gamma(T/B) \sim (T/B)^{-m}$ with $m\approx 0.46$ for $x = 0.02$ [see Figs. \ref{fig:Gamma-scaling}a and \ref{fig:Gamma-scaling}c] and $m \approx 0.5$ for $x = 0.05$ [see Figs. \ref{fig:Gamma-scaling}b and \ref{fig:Gamma-scaling}d]. 
A more accurate estimation of the location of this hidden QCP reveals $x=x_c \approx 0.03$ by extrapolating the singular specific heat coefficient in the SM state under fields as Nd doping is tuned across the transition (see Fig. \ref{fig:PD}b) \cite{Hashimoto-pnictide,Hussey-pnictide0-RPP}. When superconductivity is fully suppressed by magnetic field, a crossover from the non-Fermi liquid SM to the Fermi-liquid state with $T^2$ resistivity is clearly observed on two sides of the transition ($x=0.02$ and $x=0.05$) (see Fig. \ref{fig:rho-scaling}c). This indicates a putative quantum critical (QC) line extended from the QCP hidden under superconducting dome at zero field and Nd doping (see Fig. \ref{fig:PD}a) \cite{Thompson-PRL-2011}. 

Here,we propose a microscopic mechanism to account for the above phenomena based on the interplay between Kondo screening
of a local $f$-electron ($f_{i\sigma}$ on site $i$ with spin $\sigma$), by mobile electrons ($c_{i\sigma}$), and the AF correlations between nearest-neighbor spins of local $f$-electrons in the form of quasi-$2d$ ($d=2+\eta$ with $ 0<\eta<1$) Kondo-Heisenberg (KH) model (see \textit{Methods}). This mechanism has been used to qualitatively describe the AF QCP at $x_c$ inside the superconducting dome of CeRhIn$_5$ under pressure \cite{Thompson-NJP-CeRhIn,Thompson-Nature-CeRhIn}, which bears a striking similarity to that for Ce$_{1-x}$Nd$_x$CoIn$_5$ at zero field under Nd doping.
The Anderson's resonating-valence-bond (RVB) spin-liquid state \cite{Anderson-RVB}, defined by the spatially homogeneous  bosonic spin-singlet pair, $\Delta_{RVB} \equiv J_{H} \sum_{\sigma} \langle\text{sgn}(\sigma) f^{ \dagger}_{i\sigma}f^{ \dagger}_{j,-\sigma}\rangle$, on nearest-neighbor sites $i,\,j$, is introduced here for the AF Heisenberg term with the exchange coupling $J_H$. The Kondo hybridization is described by the spatially homogeneous  bosonic $\chi$ field where, at the mean-field level, $\chi \equiv  J_K \sum_{\sigma}\langle f^{\dagger}_{i\sigma}c_{i\sigma}\rangle$ with Kondo coupling $J_K$. 
At the edge of antiferromagnetism, the Kondo effect not only stablizes the RVB spin liquid against the magnetic long-ranged order by partially sharing the $f$-electron spins \cite{Coleman-Andrei}, but also introduces hoping of the RVB bonds to the conduction band to form charged Cooper pairs. This leads to a Kondo-RVB coexisting heavy-electron superconducting state with estimated transition temperature $T_c \sim \chi^2 \Delta_{RVB}$, in quantitatively good agreement with the experimental observation \cite{Chang-SSC-PRB}. 

Via the competition between Kondo and RVB physics, an AF-KB QCP was predicted inside the superconducting dome, qualitatively describing the phase transition between AF-superconducting coexisting phase and a pure superconducting phase observed in CeRhIn$_5$\cite{Chang-SSC-PRB}, similar to our case here. By analyzing the amplitude  fluctuations of the Kondo and RVB  correlations beyond mean-field level via renormalization group (RG) analysis\cite{Chang-PRB-SM,Chang-SSC-PRB} (see Supplementary Notes, section \ref{app:rg}), this mechanism provides a qualitative and semi-quantitative understanding of the SM properties in CeCoIn$_5$ in fields and pressure near the AF-KB QCP \cite{Petrovic-JPCM,Thompson-PRL-2011, Thompson-SC-cecoin5}. In particular, it captures the observed  $T$-linear resistivity in terms of critical Kondo (charge) fluctuations via the electron-phonon-like interaction, $  \hat{H}_K \sim J_K \sum_{i\sigma} 
\left(  c_{i \sigma}^{\dagger} f_{i \sigma}   \hat{\chi}_{i} +H.c.\right)$ (see Eq. (\ref{eq:sigma}) below and Eq. (\ref{eq:HS-fluc}) in \textit{Methods}), and a power-law-in-$T$ divergence in $\gamma(T)$  via both fluctuations.  Meanwhile, the critical Kondo coupling therein depends on the anomalous dimension $\eta$ of our quasi-$2d$ model (see below). This quasi-$2d$ nature of our theoretical framework, essential to the KB transition, is consistent with the quasi-$2d$ nature of the Fermi surface in CeCoIn$_5$  
\cite{Shishido-CeCoIn-dhva} as well as the  dimensional crossover in Fermi surface evolution of the $\alpha$-band  from $2d$-like to $3d$-like  in our system\cite{Petrovic-PRB-FS} with increasing Nd concentrations observed  in quantum oscillations. By analyzing the data of quantum oscillation in Ref. \cite{Petrovic-PRB-FS}, we observed a rapid change in charge carrier $n$ across $x_c$ as $T \to 0$ (see Table \ref{eq:table-alpha}), a possible signature of a jump in Fermi surface volume at ground state. This signature is  consistent with our KB QCP scenario. Interestingly, evidence of a delocalization (KB) transition has been observed in a closely related compound CeCo(In$_{1-x}$Sn$_x$)$_5$ near $x\sim 0.016$ \cite{Sn-115}. It is promising to expect that the KB QCP might occur here. The above  experimental development strongly motivates us to apply this theoretical framework for our system.


To account for the Planckian scattering rate, we first computed  the electrical resistivity $\rho(T)$ near the KB QCP via electron scattering rate $1/\tau$, proportional to imaginary part of conduction electron self energy, contributed from the critical Kondo fluctuations within the Boltzmann transport theory and renormalized perturbation. 
We find the perfect $T$-linear resistivity  with precise form of Kondo-coupling-dependent  $A_1$ coefficient, given by,
\begin{align}
     &\rho(T) = \rho_0 + A_1 T, \nn 
     &\rho_0 = \frac{m^\star}{n e^2} \left(\frac{ \pi}{N_0(h/2\pi)}\right), \, A_1 = \frac{m^\star}{ne^{2}}\left(\frac{Y}{\pi  j_{K}^{2}}\right)\frac{k_{B}}{(h/2\pi)}
    \label{eq:sigma}
\end{align}
where $j_K = J_K N_0$ is the dimensionless Kondo coupling with $N_0$ being the conduction electron density of states at the Fermi energy,   $j^\star_K =\sqrt{\eta}$ at the KB QCP, and the constant $Y\approx 1.39$ (see Ref. \cite{Chang-PRB-SM} and Supplementary Notes, sections \ref{app:rho} and \ref{app:rho-2}). From Eq. (\ref{eq:sigma}), we have $\alpha = Y/\pi j_K^2 $ \cite{Chang-SSC-PRB}, consistent with the relation $ \alpha  \propto 1/T_{\text{coh}} \propto 1/j_K^2$ \cite{Georges-PRL-2000} (see Supplementary Notes, section \ref{app:crossover}). At criticality, we have 
\begin{align}
    \alpha = \frac{Y}{\pi \eta}.
    \label{eq:alpha}
\end{align}
Surprisingly, though $\alpha$ depends on microscopic coupling constant, at QCP it reduces to depend only on the anomalous spatial dimension $\eta$, a macroscopic constant. By properly choosing the fitting parameter $\eta \approx 0.45$, corresponding to a critical Kondo coupling $j_K^\star \approx 0.7$, the Planckian limit $\alpha \approx 1$ is recovered. The value of $j_K^\star$ agrees reasonably well with the ARPES measurement for CeCoIn$_5$: $j_K = -\left[\ln (k_B T^*_{\text{onset}} / D)\right]^{-1} \approx 0.62$ with $D \sim 600$K and $T^*_{\text{onset}}\sim 120$K being the onset temperature for Kondo correlation\cite{Georges-PRL-2000,Yuan-115}.
The approximated constant value of $\alpha \sim O(1)$ seen for $0\leq x\leq 0.05$ (with an error bar, see Table \ref{eq:table-alpha}) can therefore be understood as the Kondo hybridization being close to  quantum critical. This is also in good agreement with the same slope observed in linear-in-$T$ resistivity near the edge of antiferromagnetism as well as  the decreasing $1/A_1$ (or Kondo correlation) with increasing Nd doping (see Fig. \ref{fig:parameter} and Supplementary Notes, section \ref{app:crossover}). The $T$-linear resistivity we theoretically obtain here can be understood as a generic feature of gapless real-valued bosons (quadratically dispersed bosonic  Kondo hybridization $\hat{\chi}$ field) coupled to a fermionic particle-hole pair (made of a conduction electron and a fermionic spinon). This interaction form is reminiscent of the electron-phonon coupling in conventional metals where $T$-linear resistivity has been known to exist in the high-temperature limit\cite{hartnoll-planckian}. However, unlike the high-temperature electron-phonon effect,  the linear-in-$T$ resistivity observed here occurs only at very low temperatures. Furthermore, the experimentally and theoretically supported relations between $A_1$  and the Kondo coupling $j_K$, $\alpha \propto A_1 \propto 1/j_K^2$, as well as $\alpha \sim 1$ strengthen the link between the Planckian $T$-linear scattering rate and critical Kondo hybridization in our system.


To further support the above theoretical analysis, we calculated the electronic specific coefficient $\gamma(T)$ near the KB QCP, dominated by the bosonic  Gaussian fluctuations of the Kondo hybridization ($\hat{\chi}$) near criticality. 
This leads to universal scaling of $\gamma(T)$:
\begin{align}
    \gamma(T) = C_V/T \sim |g-g_c|^{-\beta} \Phi(T/T_{LFL})
    \label{eq:gamma-critical}
\end{align}
with $\beta = 3\eta^2/8 + \eta/4$ (see Supplementary Notes, section \ref{supp:gamma}). Here, $\Phi(T/T_{LFL})$ is a universal scaling function and $T_{LFL} \sim |g-g_c|$ (see Supplementary Notes, section  \ref{supp:gamma}). Strikingly, Eq. (\ref{eq:gamma-critical}) agrees remarkably well with the experimental observations for $x=0.02, 0.05$ with the same $\eta=0.45$ (see dashed curves in Figs. \ref{fig:Gamma-scaling}a and \ref{fig:Gamma-scaling}b, and the Supplementary Notes, section \ref{supp:gamma}). In particular, $\Phi(T/T_{LFL})$ from theory well describes the singular $T/B$-power-law scaling behavior in $\gamma(T/B)\sim (T/B)^{-m}$ observed experimentally for $0.01<T/B<1$ with $m\sim 0.5$ (see dashed lines in Figs. \ref{fig:Gamma-scaling}c and \ref{fig:Gamma-scaling}d). Furthermore, we find the predicted critical exponents $\beta \sim 0.16$ in Eq. (\ref{eq:gamma-critical}) is  in excellent agreement with the asymptotically power-law behavior of the normal-state electronic specific heat coefficient as a function of $|x-x_c|\propto |g-g_c|$ at low temperatures  where $\gamma(T\to0.1K) \sim |x-x_c|^{-\beta}$ with $\beta \approx 0.1$ for $x<x_c$ and $\beta \approx 0.16$ for $x>x_c$, see the pink curves of Fig. \ref{fig:PD}b. Note, however, that the above quantum critical features are inconsistent with the conventional spin-density-wave (SDW) theory for the AF QCP though the SDW fluctuations are expected to appear\cite{Lohneysen-RMP}. 

A few remarks are made before we conclude. Firstly, since $m^\star$ is likely temperature dependent in the SM region, Planckian scattering rate is reached only near the lower end of the linear-in-$T$ resistivity region where $m^\star$ is well-approximated by a temperature-independent constant extracted from quantum oscillation measurement in the Fermi-liquid region \cite{hartnoll-planckian}. Secondly, though the Drude formula has been widely used to relate the transport and scattering rate in the Fermi-liquid region and near the lower end of the $T$-linear region, whether it is valid  deep in the non-Fermi liquid SM region is an open question \cite{hartnoll-planckian,paschen-arxiv-2022}. Finally,  while quantum oscillation measurement is widely used to estimate $n$ and $m^\star$, different approach by Hall coefficient and $A$ coefficient associated with the Fermi-liquid behavior with $T^2$ resistivity has been used \cite{paschen-arxiv-2022}.

In summary, we observe the quantum-critical scaling of the Planckian metal state in the resistivity and specific heat coefficient of heavy-electron superconductor Ce$_{1-x}$Nd$_x$CoIn$_5$ under fields.
Clear experimental evidences are shown to support the notion of Kondo hybridization being quantum critical at $x_c \sim 0.03$. 
Our proposed mechanism based on the critical Kondo fluctuations simultaneously describes remarkably well the observed universal $T$-linear Planckian scattering rate with $\alpha \sim 1$, the quantum critical scaling in resistivity and specific heat coefficient as well as the power-law divergence in specific heat coefficient. The $\alpha$ coefficient is found to in general depend microscopically on the Kondo hybridization. Surprisingly, however  at criticality, it depends only on the quasi-$2d$ nature of the system and reaches the universal Planckian limit. Our observation and proposed mechanism offers the first microscopic understanding of the Planckian dissipation limit in a quantum critical system.
This generic mechanism is relevant for other heavy-fermion quantum critical superconductors showing Planckian metal states, such as CeRhIn$_5$ and related compounds. Our results motivate further experimental study, such as the Hall coefficient measurement on the signatures of  Kondo breakdown near the critical Nd doping as well as the fundamental issue on whether quantum critical systems with $T$-linear resistivity, in general, implies the Planckain metals.



\section*{Methods}
\textbf{A microscopic model.} Our start point is the microscopic large-$N$ (Sp($N$)) Kondo-Heisenberg Hamiltonian  $H = H_0 +  H_K + H_J$, where  
\begin{align}
	& H_{0}=  \sum_{\langle i, j \rangle; \sigma}\Big[ t_{ij}c^{\dagger}_{i\sigma}c_{j\sigma}  +  H.c.  \Big]  - \sum_{i\sigma} \, \mu \, c_{i\sigma}^{\dagger} c_{i\sigma}, \nn 
	 &  H_{J} =  \sum_{\langle i,j \rangle}\frac{J_{ij}}{N}\mbd{S}_{i}^{f} \cdot \mbd{S}_{j}^{f}, \quad H_{K} =  \frac{J_{K}}{N} \sum_{i} \mbd{S}_{i}^{f} \cdot \mbd{s}^{c}.
\label{eq:H}
 \end{align}
In Eq. (\ref{eq:H}), $H_0$ describes hopping of conduction ($c_{i\sigma}$) electrons. The antiferromagnetic Heisenberg interaction between two impurity-spin ($\mbd{S}^f$) of the electrons occupying the localized $4f$ orbitals of neighboring sites is described by  $H_J$ and the Kondo screening of impurity spins by conduction electrons is captured by $H_K$. Via Hubbard-Stratonovich transformation, $H_K$ and $H_J$ can be factorized as
\begin{align}
    & H_{J} \to  \sum_{\langle i,j \rangle;\alpha, \beta } \left[ \Phi_{ij}\mathcal{J}^{\alpha \beta} f_{i\alpha}f_{j\beta}+H.c. \right]  + \sum_{\langle i,j \rangle}N \frac{|\Phi_{ij}|^{2}}{J_{H}}~,\nn
      & H_{K} \to   \frac{1}{\sqrt{N}}\sum_{i,~\sigma} 
\left[ \left( c_{i \sigma}^{\dagger} f_{i \sigma}  \right) \chi_{i} +H.c. \right]  + \sum_{i}\, \frac{|\chi_{i} |^{2}}{J_K}.
\label{eq:HS}
\end{align}
Here, the local spin operator $\mbd{S}^f_i$ is represented in terms of the constraint fermionic Sp($N$) fields, $f_{i\alpha}$, which needs to be subjected to the constraint, $\langle \sum_\sigma f_{i\sigma}^{\dagger}f_{i \sigma} \rangle = N\kappa$, to ensure its local nature. The constant $\kappa <1$ allows us to capture the valence fluctuations. Thus, to capture the local constraint, an additional term has to be included in the Hamiltonian, $H_{\lambda}  =   \sum_{i, \sigma}\lambda \Big[f_{i\sigma}^{\dagger}f_{i \sigma}-Q  \Big] $ with $\lambda$ being the Lagrange multiplier. 
In the $H_J$ term, we assume an uniform antiferromagnetic RKKY coupling $J_{ij}=J_H$ on a lattice where $i,~j$ are nearest-neighbor site indices, $\mathcal{J}^{\alpha \beta}=\mathcal{J}_{ \alpha\beta}=-\mathcal{J}^{\beta\alpha }$ denotes the anti-symmetric tensor,  $\sigma,\alpha,\beta \in \lbrace -\frac{N}{2}, \cdots, \frac{N}{2} \rbrace$ label the spin indices. Here, $\chi_i$ and $\Phi_{ij}$ are the spatially dependent Hubbard-Stratonovich fields, where their (spatially uniform) mean-field values, $ \chi \equiv \langle \frac{J_K}{\sqrt{N}}\sum_{\sigma} f^{\dagger}_{i\sigma}c_{i\sigma}\rangle$ and $ \Delta_{RVB} \equiv \langle \Phi_{ij} \rangle  = \langle \frac{J_{H}}{N} \sum_{\alpha, \beta} \, \mathcal{J}_{\alpha \beta}f^{\alpha \dagger}_{i}f^{\beta \dagger}_{j}\rangle$, can be used as order parameters to describe the Kondo-screened heavy Fermi liquid state and the RVB spin-liquid state. 

While considering the fluctuations of $\chi_i$ and $\Phi_{ij}$ beyond mean-field level, we can express $\chi_i \to \chi+J_K \hat{\chi}_{i}$ and  $\Phi_{ij}\to \Delta_{RVB} + J_H \hat{\Phi}_{ij}$, where $\hat{\chi}_i$ and $\hat{\Phi}_{ij}$ denote the fluctuation of $\chi$ and $\Delta_{RVB}$. Both $H_K$ and $H_J$ can be divided into the mean-field and fluctuating parts, namely $H_K \to H^{MF}_K + \hat{H}_K$ and $H_J \to H^{MF}_J + \hat{H}_J$, with 
\begin{align}
    & H_{J}^{MF} = \Delta_{RVB} \sum_{\langle ij \rangle;\alpha \beta } \left[ \mathcal{J}^{\alpha \beta} f_{i\alpha}f_{j\beta}+H.c. \right]  + 4 N N_s \frac{\Delta_{RVB}^{2}}{J_{H}}~,\nn
      & H_{K}^{MF} =   \frac{ \chi}{\sqrt{N}}\sum_{i\sigma} 
\left[  c_{i \sigma}^{\dagger} f_{i \sigma}   +H.c. \right]  +  \frac{ N \chi^{2}}{J_K},
\label{eq:HS-mf}
\end{align}
at mean-field, and 
\begin{align}
    &\hat{H}_{J} =  J_H \sum_{\langle ij \rangle;\alpha \beta } \left( \hat{\Phi}_{ij}\mathcal{J}^{\alpha \beta} f_{i\alpha}f_{j\beta}+H.c. \right)  + \sum_{\langle i,j \rangle}N J_H |\hat{\Phi}_{ij}|^{2}~,\nn
      & \hat{H}_{K} =    \frac{J_K}{\sqrt{N}}\sum_{i\sigma} 
\left(  c_{i \sigma}^{\dagger} f_{i \sigma}   \hat{\chi}_{i} +H.c. \right)  + \sum_{i}\, J_K |\hat{\chi}_{i} |^{2}
\label{eq:HS-fluc}
\end{align}
beyond mean-field level. 

The microscopic model relevant for the heavy-fermion materials therefore becomes $H\to H_0+H_\lambda + H_J^{MF} +H_K^{MF} + \hat{H}_K + \hat{H}_J$.

\section*{Acknowledgements}
We acknowledge discussions with S. Kirchner, C.-L. Huang , J. Thompson, A. P. Mackenzie and S. A. Hartnoll. This work is supported by the Ministry of Science and Technology Grants 107-2112-M-009-010-MY3, the National Center for Theoretical Sciences of Taiwan, Republic of China (to C.-H.C.). Work at BNL is supported by the Office of Basic Energy Sciences, Materials Sciences and Engineering Division, U.S. Department of Energy (DOE) under Contract No. DESC0012704 (materials synthesis, thermodynamic and transport characterization). 
\section*{Data availability}
All relevant data are available from the authors upon reasonable request.
\section*{Additional Information}
\subsection*{Author Contributions} All authors contributed to the theoretical research described in the paper and the writing of the manuscript.
\subsection*{Competing financial interests}
The authors declare no competing financial or non-financial interests.


\setcounter{equation}{0}
\setcounter{figure}{0}
\renewcommand{\theequation}{S\arabic{equation}}
\renewcommand\thesection{S.\Roman{section}}
\renewcommand{\figurename}{Fig. S}

\section{Supplementary Notes: Specific heat coefficient and its scaling for zero Nd doping ($x=0$)}
\begin{figure*}[ht]
     \centering
     \includegraphics[width=0.9\textwidth]{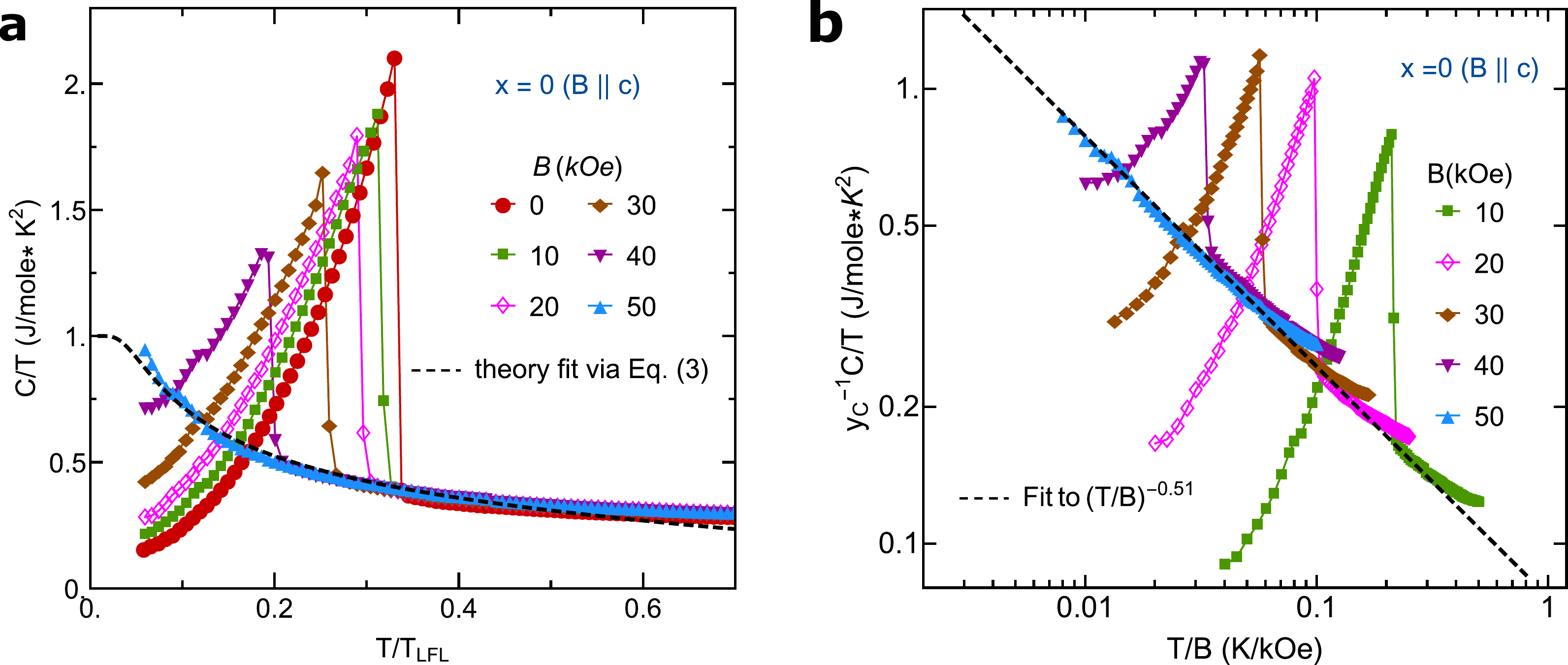}
     \caption{\textbf{Specific heat coefficient $C_V/T$ and its scaling for zero Nd doping ($x=0$).} \textbf{a} shows the electronic specific heat coefficient $C_V/T$ with different fields $B \parallel c$ for zero Nd doping ($x=0$) while \textbf{b} displays the power-law $T/B$ scaling of \textbf{a}. }
     \label{fig:zeroNd-gamma}
 \end{figure*}

Fig. S\ref{fig:zeroNd-gamma} shows the specific heat coefficient and its scaling at zero Nd doping ($x=0$) under different magnetic fields. The experimentally observed specific heat coefficient for zero doping in the non-Fermi liquid regime ($B = 50$kOe)  agrees well with Eq. (\ref{eq:gamma-critical}) (see Fig . S\ref{fig:zeroNd-gamma}a), similar to the cases of $x = 0.02$ and $x= 0.05$.  A $T/B$-power-law scaling behavior within the quantum-critical regime, $\gamma(T/B) \sim (T/B)^{-m}$, with exponent $m = 0.51$, closed to the exponents for $x = 0.02$ and $x= 0.05$, is also found here (Fig. S\ref{fig:zeroNd-gamma}b).
\label{app:zeroNd}

\section{Supplementary Notes: Crossovers scales and the $A_1$ coefficient}
\label{app:crossover}
\begin{figure}[t]
     \centering
     \includegraphics[width=0.45\textwidth]{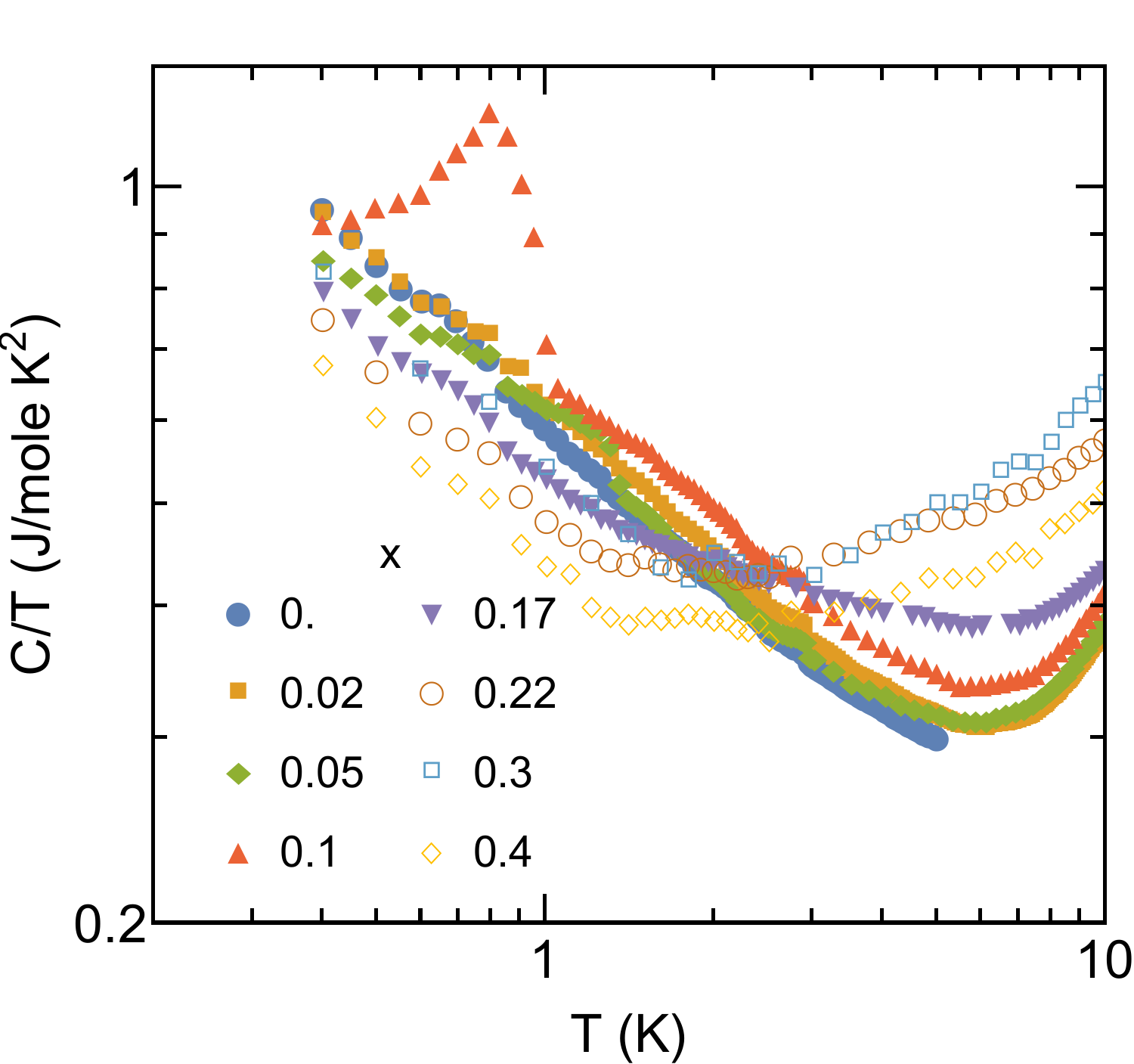}
     \caption{\textbf{Specific heat coefficient $C_V/T$ at a finite field.} Specific heat coefficient at a finite field ($B = 70$kOe along the $c$ axis) with different Nd concentrations ($x$). Note that $C_V/T$ for $x=0$ is measured under $B = 50$ kOe.}
     \label{fig:gamma-HiB}
 \end{figure} 
Based on the theory in Ref. \cite{Georges-PRL-2000}, the coherent temperature $T_{\text{coh}}$ is related to the condensation amplitude of the Kondo hybridization field $\chi$ through $T_{\text{coh}}\sim \chi^2/D$ with $D$ being the conduction-electron bandwidth. As the amplitude of the Kondo hybridization field also proportional to the Kondo coupling, $\chi \propto J_K$, this suggests 
 \begin{align}
     T_{\text{coh}}\sim J_K^2/D.
     \label{eq:Tcoh-Jk}
 \end{align}
In the previous papers \cite{Coleman-frustration-KL,Chang-SSC-PRB}, the authors demonstrated that the superconducting transition temperature of a heavy-fermion superconductor can be estimated as
\begin{align}
    T_c \sim \chi^2 \Delta_{\text{RVB}} \sim J_K^2 \Delta_{\text{RVB}}
    \label{eq:Tc-Jk}
\end{align}
with $\Delta_{\text{RVB}}$ being the order parameter of the RVB spin-singlet bond. Hence, Eqs. (\ref{eq:Tcoh-Jk}) and (\ref{eq:Tc-Jk}) implies the slope of the $T$-linear resistivity for Ce$_{1-x}$Nd$_x$CoIn$_5$ for $x$ less than 0.17 is directly related to the Kondo coupling through
\begin{align}
    1/A_1 \sim J_K^2.
    \label{eq:A1-JK}
\end{align}

\section{Supplementary Notes: Estimation of carrier concentration $n$ and $\alpha$ coefficient}
\label{supp:esti-alpha}

In this section, we provide  derivation of the relevant equations for carrier concentration $n$ based on the quantum oscillation measurements.  We will further use those equations to reproduce $n$ and the Planckian coefficients $\alpha$ shown in Table \ref{eq:table-alpha} of the main text.

We start from the formula of  quantum oscillation frequency $F$, given by
\begin{align}
    F=\frac{(h/2\pi)}{2\pi e}S_{F}
\end{align}
with $S_{F}$ being the extremal cross-sectional area of the Fermi surface. For simplicity, we assume a circular cross-section of Fermi surface here, hence $ S_{F}=\pi k_{F}^{2}$ with $k_{F}$ being the ``averaged'' Fermi wave vector of the circular Fermi surface. This links the dHvA frequency and the ``average'' Fermi wave vector of the Fermi surface by  $F=\frac{(h/2 \pi) k_{F}^{2}}{2e}$. Here, we can make a link of $F$ and the carrier concentration $n$ through $k_F$. While considering the effective dimensionality of critical modes, the carrier concentration takes the following form\cite{Mackenzie-Science}:
\begin{align}
    & n = \frac{2k_F}{\pi d_b d_c} = \frac{2}{\pi d_b d_c} \sqrt{\frac{2eF}{h/2\pi}} \quad (\text{for 1d}),\nn
    & n = \frac{k_F^2}{2\pi d_c} = \frac{1}{2\pi d_c} \left( \frac{2eF}{h/2\pi}\right) \quad (\text{for 2d}),\nn
    & n = \frac{k_F^3}{3\pi^2} = \frac{1}{3\pi^2} \left( \frac{2eF}{h/2\pi}\right)^{3/2} \quad (\text{for 3d}),
    \label{eq:n-mackenzie}
\end{align}
where $d_b$ and $d_c$ are the lattice constants of unit cell along the $b$ and $c$ axes. Note that, for the $2d$ case of Eq. (\ref{eq:n-mackenzie}), we assume the system has a strong anisotropy along the $c$ direction while it remains isotropic in the a-b plane.

As an example, we provide detailed derivation of carrier concentration for the $2d$ case shown in Eq. (\ref{eq:n-mackenzie}) and then generalize this derivation to the case with  fractional quasi-$2d$ dimension.

Assume the critical modes occur on the isotropic $ab$ plane. The total number of states can be expressed as
\begin{align}
   & N	=N_{ab}N_{c}=2\times\frac{\pi k_{F}^{2}}{\Delta V_{k}}\times\frac{L_{c}}{d_c}=2\times\frac{A\pi k_{F}^{2}}{4\pi^{2}}\frac{L_{c}}{d_c} \nn
    & \to n=\frac{N}{\mathcal{V}}=\frac{N}{AL_{c}}=\frac{k_{F}^{2}}{2\pi d_c},
\end{align}    
where $N_{ab}=\frac{\pi k_{F}^{2}}{\Delta V_{k}}$ is the number of states on the $ab$ plane per spin while $N_{c}=L_{c}/d_c$ for that along the $c$ axis. Here, $\Delta V_{k}$ represents the unit volume in $k$ space occupied by a state, $L_{c}$ denotes the sample site along $c$, $A=L^{2}$ is the area of the sample on the $ab$ plane. This indicates the total volume of the sample $\mathcal{V}=AL_{c}$.

The above approach of deriving the carrier concentration for the effective $2d$ critical modes embedded in a $3d$ lattice can be generalized for the case of arbitrary dimensional critical modes with fractional quasi-$2d$ dimensionality embedded in a $3d$ lattice. 

The total number of states per spin for an isotropic $d$-dimensional system is given by 
\begin{align}
    N=\sum_{|\bm{k}|\leq k_{F}}\Theta(\varepsilon_{\bm{k}})=\frac{\int_{0}^{k_{F}}d^{d}k}{\Delta V_{k}}=\frac{V_{d}(k_F)}{\Delta V_{k}},
    \label{eq:N-d-dim}
\end{align}
where $\Delta V_{k}=(2\pi/L)^{d}= (2\pi)^d/V_d$ with $V_d \equiv L^d$ and $V_{d}(k_F)$ is the volume of a $d$-dimensional sphere with radius $k_F$,
\begin{align}
    V_{d}(k_F)=\frac{\pi^{d/2}}{\Gamma(\text{\ensuremath{\frac{d}{2}+1}})}k_F^{d}.
\end{align}
In the above equation, $\Gamma (x)$ denotes the $\Gamma$ function.

For a general $d$-dimensional critical modes embedded in a $3d$ lattice, the total number of states reads
\begin{align}
    N=2\times N_{ab}^{(d)}\times N_{c}^{(3-d)},
\end{align}
where the prefactor $2$ comes from the spin degrees of freedom. Here, we assume that the quasi-$2d$ critical modes mostly arise from the $ab$-plane.

From Eq. (\ref{eq:N-d-dim}), we have
\begin{align}
    N^{(d)}_{ab}=\frac{\frac{\pi^{d/2}}{\Gamma(\text{\ensuremath{\frac{d}{2}+1}})}k_{F}^{d}}{\frac{(2\pi)^{d}}{V_{d}}}=\frac{\pi^{d/2}V_{d}k_{F}^{d}}{(2\pi)^{d}\times\Gamma(\text{\ensuremath{\frac{d}{2}+1}})}, 
\end{align}
while
\begin{align}
    N^{(3-d)}_{c}=\left(\frac{L_{c}}{d_c}\right)^{3-d}
\end{align}
The total number of states is then given by
\begin{align}
    N=\frac{\mathcal{V}k_{F}^{d}}{2^{d-1}\pi^{d/2}d_c^{3-d}\Gamma(\text{\ensuremath{\frac{d}{2}+1}}) },
\end{align}
giving rise to the carrier concentration 
\begin{align}
    n=\frac{N}{\mathcal{V}}=\frac{k_{F}^{d}}{2^{d-1}\pi^{d/2}d_c^{3-d}\Gamma(\text{\ensuremath{\frac{d}{2}+1}})}.
\end{align}
Using the relation of $k_F$ and $F$, we obtain the expression of carrier concentration for arbitrary $d$-dimensional critical modes,  
\begin{align}
    n=\frac{1}{2^{d-1}\pi^{d/2}d_c^{3-d}\Gamma(\text{\ensuremath{\frac{d}{2}+1}})}\left(\frac{2eF}{h/2  \pi}\right)^{\frac{d}{2}}.
    \label{eq:n-fractional-d}
\end{align}
When taking $d=2$, the above expression of $n$ goes back the $2d$ case in Eq. (\ref{eq:n-mackenzie}).

\subsection{Estimating the Planckian  coefficients $\alpha$ for Ce$_{1-x}$Nd$_x$CoIn$_5$}
Below, we estimate the carrier concentration $n$ and Planckian  coefficients $\alpha$ shown in Table \ref{eq:table-alpha} for Ce$_{1-x}$Nd$_x$CoIn$_5$ with $x = 0,\, 0.02, \, 0.05$, and $0.1$ using  the dHvA frequency $F$ and effective mass $m^\star$ in Ref. \cite{Petrovic-PRB-FS} (for $\alpha$-band) and in Refs. \cite{Shishido-CeCoIn-dhva,Mackenzie-Science} (for $\beta$-band).

\begin{itemize}
    \item \textbf{For} $\bm{x = 0}$. The average dHvA frequency for the $\alpha$-band is $F=4.9$kT while $m^\star = 11.7m_{0}$ is  its average effective mass. Since Fermi surface of pure CeCoIn$_5$ has been shown to be $2d$-like, we thus use the equation of the $2d$ version in Eq. (\ref{eq:n-mackenzie}) to estimate the carrier density, given by
    \begin{align}
        n=& \frac{\left(2\times1.6\times10^{-19}\right)\times(4.9\times10^{3})}{(7.549\times10^{-10})\times\left(6.6\times10^{-34}\right)}\text{m}^{-3}\nn
        & =0.31\times10^{28}\text{m}^{-3}\quad(\text{for }\alpha\text{-band}).
    \end{align}
For the $\beta$-band of pure CeCoIn$_5$, we adopt its Fermi surface parameters:  $F=9.75$kT and $m^\star = 100m_{0}$. Following the similar approach, the carrier concentration for $\beta$-band is
\begin{align}
    n = 0.63\times 10^{28}\text{m}^{-3} \quad (\text{for }\beta \text{-band}).
\end{align}

The Planckian coefficient from the $\alpha$-band can be straightforwardly obtained via $\alpha=\frac{e^{2}(h/2\pi)}{k_{B}}A_{1}\frac{n}{m^\star}$, suggesting
     \begin{align}
         \alpha =& \frac{\left(1.6\times10^{-19}\right)^{2}\cdot \left(0.98\times10^{-8}\right)\left(0.31\times10^{28}\right)}{\left(1.38\times10^{-23}\right)\cdot \left(11.7\times 9.11\times10^{-31}\right)} \nn
         & \quad\quad\times \left(1.05\times10^{-34}\right) \nn 
         = & 0.56\quad(\text{for }\alpha\text{-band}),
     \end{align}
     and, for the $\beta$-band, we have
     \begin{align}
          \alpha = 0.13\quad(\text{for }\beta\text{-band}).
     \end{align}
    These two contributions give $\alpha \approx 0.7$ for pure CeCoIn$_{5}$. The $A_{1}$ coefficient for pure CeCoIn$_{5}$ at zero field, $A_{1}=0.98\mu\Omega\cdot\text{cm}/\text{K},$ is used for the estimation of the $\alpha$ coefficients shown above.
    \item \textbf{For} $\bm{x = 0.02}$.  The average dHvA frequency for the $\alpha$-band is $F=4.89$kT while $m^\star = 11.7m_{0}$ is its average effective mass. Angular dependence of the dHvA frequencies indicates a $2d$ Fermi surface of the $\alpha$-band for $x = 0.02$\cite{Petrovic-PRB-FS}.  Following the similar approach, the carrier concentration of the $\alpha$-band is calculated as
    \begin{align}
        n = 0.32\times 10^{28}\text{m}^{-3}\quad(\text{for }\alpha\text{-band}).
    \end{align}
    Here, we assume that the carrier density and the relevant band parameters as well as the effective dimension of the $\beta$-band do not significantly altered while doping $2\%$ of Nd, indicating that $n = 0.63\times10^{28}\text{m}^{-3}$ and $m^\star = 100m_0$ for the $\beta$-band here. The $\alpha$-coefficients for the $\alpha$- and $\beta$-band are thus estimated 
    \begin{align}
        &\alpha = 0.58\quad(\text{for }\alpha\text{-band}), \nn
        & \alpha = 0.14\quad(\text{for }\beta\text{-band}),
    \end{align}
    giving the total Planckian coefficient $\alpha = 0.72$ for $x = 0.02$. The gradient of the linear-$T$ resistivity $A_1 = 1.0\mu \Omega\cdot \text{cm}/K$ is used in this case. 
    \item \textbf{For} $\bm{x = 0.05}$. The fundamental band parameters of the $\alpha$-band for $x = 0.05$ is $F=4.88$kT and $m^\star = 9.15m_{0}$. Angular dependence of the dHvA frequencies indicates a $2d$-to-$3d$ dimensional crossover of Fermi surface of the $\alpha$-band at $x = 0.05$\cite{Petrovic-PRB-FS}. Accompanying with the prediction of a QCP at $x_c = 0.03$ and the theoretical studies on that QCP, we treat the dimensionality for $x = 0.05$ to be $d = 2.45$. Using Eq. (\ref{eq:n-fractional-d}), the carrier concentration of $\alpha$-band is estimated as
    \begin{align}
        n = 0.26\times 10^{28}\text{m}^{-3}\quad(\text{for }\alpha\text{-band}).
    \end{align}
    Likewise, we assume the band parameters of the $\beta$-band also remains the same for $x = 0.05$, thus $m^\star = 100m_0$ and $F = 9.75$kT. The carrier concentration of the $\beta$-band with $d=2.45$ is found to be
    \begin{align}
        n = 0.6\times 10^{28}\text{m}^{-3}\quad(\text{for }\beta\text{-band}).
    \end{align}
    Using  $A_1 = 1.17 \mu \Omega\cdot \text{cm}/K$ for $x = 0.05$, the $\alpha$-coefficients for the $\alpha$- and $\beta$-band can be straightforwardly calculated as
    \begin{align}
        &\alpha = 0.7\quad(\text{for }\alpha\text{-band}), \nn
        & \alpha = 0.15\quad(\text{for }\beta\text{-band}),
    \end{align}
    giving the total Planckian coefficient $\alpha = 0.85$.
    \item \textbf{For} $\bm{x = 0.1}$. The fundamental band parameters of the $\alpha$-band for $x = 0.1$ is $F=4.41$kT and $m^\star = 7m_{0}$. We treat the effective dimensionality of Fermi surface of the $\alpha$-band to be three-dimensional as this compound at $x = 0.1$ is deep inside the AF state \cite{Petrovic-PBR-NdCeCoIn}. Using Eq. (\ref{eq:n-fractional-d}) and $d = 3$, the carrier concentration of $\alpha$-band is estimated as
    \begin{align}
        n = 0.17\times 10^{28}\text{m}^{-3}\quad(\text{for }\alpha\text{-band}).
    \end{align}
    Similarly, we assume that $m^\star = 100m_0$ and $F = 9.75$kT are also applicable for the $\beta$-band for $x = 0.1$ here. The carrier concentration of the $\beta$-band with $d=3$ is found to be
    \begin{align}
        n = 0.55\times 10^{28}\text{m}^{-3}\quad(\text{for }\beta\text{-band}).
    \end{align}
    Using  $A_1 = 1.49 \mu \Omega\cdot \text{cm}/K$ for $x = 0.1$, the $\alpha$-coefficients for the $\alpha$- and $\beta$-band can be straightforwardly estimated as
    \begin{align}
        &\alpha = 0.77\quad(\text{for }\alpha\text{-band}), \nn
        & \alpha = 0.19\quad(\text{for }\beta\text{-band}),
    \end{align}
    giving the total Planckian coefficient $\alpha = 0.96$.
\end{itemize}

\section{Supplementary Notes: Renormalization group analysis for strange superconductivity in CeCoIn$_5$}
\label{app:rg}


In this section, we summarize the results of the diagrammatic renormalization group (RG) analysis of the quasi-2D Kondo-Heisenberg lattice model with the inclusion of superconducting fluctuations based on Ref.\cite{Chang-SSC-PRB}. 

The superconducting fluctuations can be effectively captured by adding the following terms in Eq. (\ref{eq:H})\cite{Coleman-Andrei,Chang-SSC-PRB},
\begin{align}
    H_{sc} = \sum_{\langle i, j\rangle}\sum_{\alpha,\beta} (\chi_i\chi_j\Phi_{ij}\mathcal{J}^{\alpha \beta} c_{i\alpha}c_{j\beta}+H.c.).
    \label{eq:H-sc}
\end{align}
Below, we address how superconductivity emerges (disappears) around the two phase boundaries: the transition of the RVB spin-liquid to coexisting superconducting phases (denoted as $g_{c1}$) and the transition of the coexisting superconducting  to the Kondo-screened Fermi-liquid phases  (denoted as $g_{c2}$). 
 
The transition of near $g_{c1}$ occurs in the ground state of the RVB spin-liquid phase ($\Delta_{RVB} \neq 0$ and $\chi = 0 $). As a result, the leading superconducting fluctuations are dominated by the fluctuations of Kondo hybridization $\hat{{\chi}}_i$ field with the RVB correlation being treated at the mean-field level, that is to neglect the RVB fluctuations. Near $g_{c1}$, we included the approximate $H_{sc}$ of  Eq. (\ref{eq:H-sc}) shown below, with an estimated bare superconducting fluctuation coupling constant $v_{sc}$ being $v_{sc} \sim J^2_K \Delta_{RVB}$
\begin{align}
     H_{sc} \to v_{sc}\sum_{\langle i, j\rangle}\sum_{\alpha,\beta} (\hat{\chi}_i \hat{\chi}_j \mathcal{J}^{\alpha \beta} c_{i\alpha}c_{j\beta}+H.c.).
     \label{eq:H-sc-gc1}
\end{align}
  
  The RG $\beta$ functions for weak coupling $J_K$ and $v_{sc}$ near $g_{c1}$ read (see \textit{Methods} and  Ref.\cite{Chang-SSC-PRB} for details)
    \begin{align}
            & j_K^\prime =\left( -\frac{\eta}{2} \right) j_{K}+\frac{1}{2} j_{K}^3, \nn
           & v_{sc}^\prime=-\eta v_{sc} + j_{K}^{2}v_{sc} + 12 v_{sc}^3,
           \label{eq:RGgc1}
        \end{align} 
where  $j_K^\prime \equiv dj_K/dl$ with $dl \equiv -d \text{ln} \Lambda$ and $\Lambda$ being the running energy cutoff. In Eq. (\ref{eq:RGgc1}), a positive (negative) coefficient implies a relevant (irrelevant) term. We find three non-trivial fixed points in Eq. (\ref{eq:RGgc1}): $\left(j^{*^2}_{K}, ~ v^{*^2}_{sc} \right) =\left( 0, ~ 0 \right),~ \left( \eta, ~ 0  \right)$ and $\left( 0, ~ \eta/12 \right)$.  The RG flow of Eq. (\ref{eq:RGgc1}) qualitatively captures the salient features of the phase diagrams of CeCoIn$_5$ and CeRhIn$_5$. See Ref. \cite{Chang-SSC-PRB} for details.

Close to $g_{c2}$, the transition occurs deeply in the Kondo phase where the mean-field Kondo correlations $\chi$ are finite and the RVB mean-field value $\Delta_{RVB}$ is negligibly small. The transition is therefore driven by the RVB fluctuation field  $\hat{{\Phi}}_{ij}$ field, and can be described by the following approximated $H_{sc}$ term near $g_{c2}$, 
\begin{align}
    H_{sc} \approx v_{sc} \sum_{\langle i, j\rangle}\sum_{\alpha,\beta} (\hat{\Phi}_{ij}\mathcal{J}^{\alpha \beta} c_{i\alpha}c_{j\beta}+H.c.)
    \label{eq:H-sc-gc2}
\end{align}
with $v_{sc}\sim \chi^2 J_H$. In the weak coupling $J_H$ and $v_{sc}$ regime, the RG equations for the transition near $g_{c2}$ read\cite{Chang-SSC-PRB}
\begin{align}
	&j_H^\prime=-\frac{d}{2} j_{H}+4j_{H}^{3}+v_{sc}^{2}j_{H},\nn 
           &v_{sc}^\prime=\left(z-\frac{d}{2} \right) v_{sc}+v_{sc}^{3}~ .
           \label{eq:RGgc2}
\end{align} 
Detailed derivations of Eq. (\ref{eq:RGgc2}) can be referred to Ref.\cite{Chang-SSC-PRB}. 
We finds a non-trivial fixed point at  $(j^{*2}_{H}, ~v_{sc}^{*2}) = (d/8,~0)$. The RG flows of Eq. (\ref{eq:RGgc2}) reveal that both $v_{sc}$ and $j_H$ grow to $(\infty,~\infty)$, corresponding to the (mean-field) coexisting Kondo-RVB superconducting phase (see Ref. \cite{Chang-SSC-PRB}).

\begin{figure}[t]
    \centering
    \includegraphics[width=0.45\textwidth]{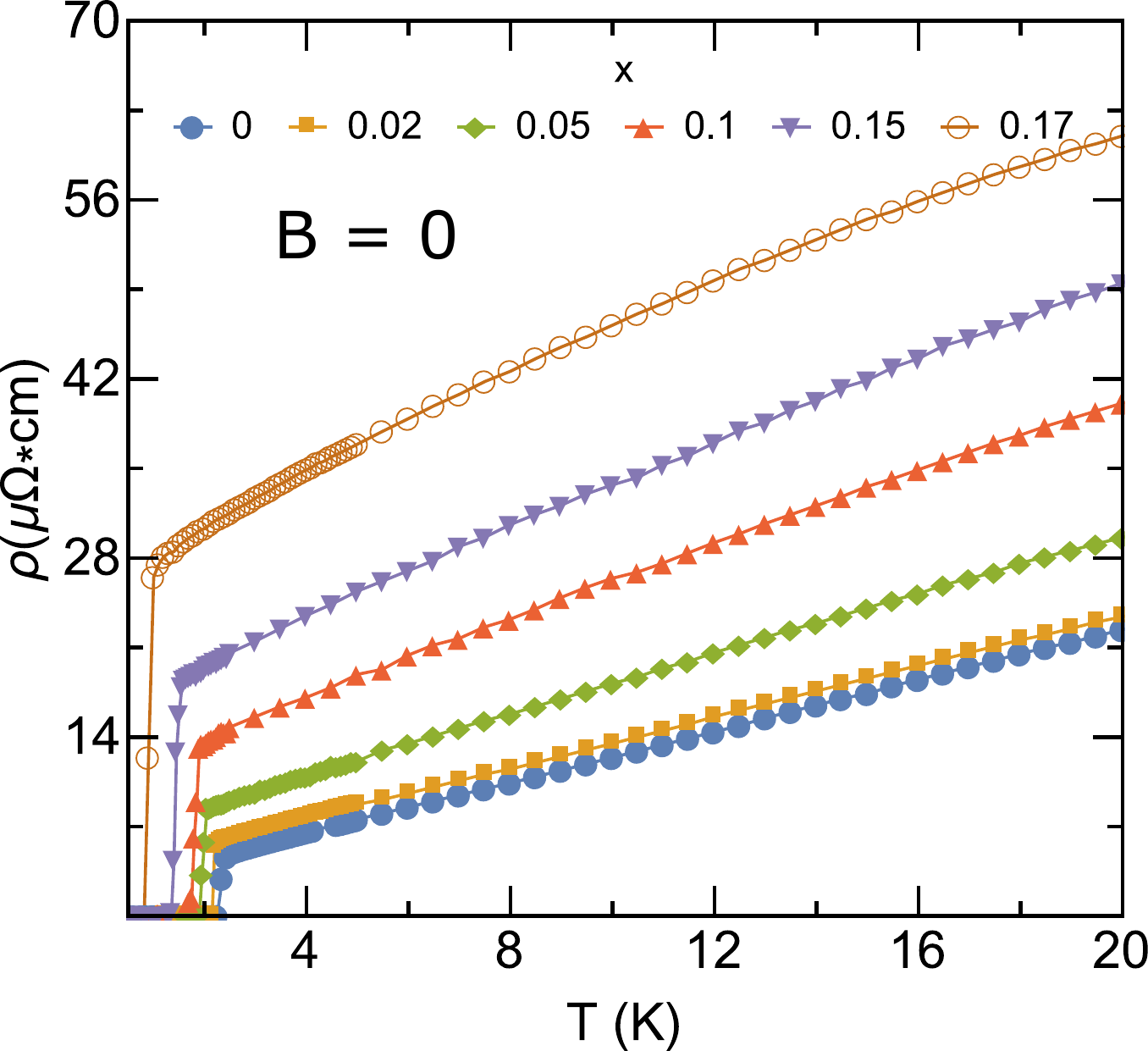}
    \caption{\textbf{Electrical resistivity under different values of Nd doping $x$ at zero field.}  The zero-field electrical resistivity of Ce$_{1-x}$Nd$_x$CoIn$_5$ with various Nd dopings. 
    }
    \label{fig:fig1}
\end{figure}

\section{Supplementary Notes: Derivation of the electrical resistivity}
\label{app:rho}
In this section, we summarize the derivation of the $T$-linear electrical resistivity based on Ref.\cite{Chang-PRB-SM} via perturbation with respect to the fluctuating Kondo Hamiltonian $\hat{H}_K$. 


The electrical resistivity $\rho(T)$ is obtained via the Boltzmann transport theory, taking the following form:
\begin{align}
    \sigma(T)=\rho^{-1}(T)=\left(-\frac{ne^{2}}{m^\star}\right)\int\tau(\omega)\frac{\partial n_{F}(\omega)}{\partial\omega}d\omega,
    \label{eq:conductivity}
\end{align}
where the relaxation time $\tau$ is directly related to the imaginary part of the conduction electron self-energy $\text{Im}\Sigma_c$, defined as
\begin{align}
    \tau(\omega)=-\frac{(h/2\pi)}{2\text{Im}\Sigma_{c}(\omega)}.
\end{align}

The imaginary part of conduction electron self-energy up to a one-loop order perturbation in $\hat{H}_K$ is given by
\begin{align}
    \text{Im}\Sigma_c (\omega) \approx \text{sgn}(\omega) \left( \Omega+\varsigma\omega \right)
    \label{eq:Sigma-c-greater}
\end{align}
with $\omega$ being measured relative to Fermi energy,  $\Omega \equiv\frac{\Lambda\pi}{2N}$ and $\varsigma\equiv\frac{1}{2\pi N j_{K}^{2}}$. 
While deriving Eq. (\ref{eq:Sigma-c-greater}), we further assume that $\Lambda/\Gamma \gg 1$ and $\omega/\Gamma \ll 1$ (or equivalently, $j_K^2 \ll 1$ and $\omega/\Lambda \ll j_K^2$). See Ref. \cite{Chang-PRB-SM} for details. 

\subsection{Derivation of electrical resistivity}
\label{app:rho-2}

Plugging the result of $\text{Im}\Sigma_c(\omega)$ into Eq. (\ref{eq:conductivity}), electrical conductivity due to fluctuating Kondo scattering can be obtained as
\begin{align}
    \sigma(T)=& \left(-\frac{ne^{2}(h/2\pi)}{2m^\star}\right)\left[\int_{0}^{\infty}\left(\frac{1}{\Omega+\varsigma\omega}\right)\frac{\partial n_{F}(\omega)}{\partial\omega}d\omega \right. \nn
   &\quad \quad \left. +\int_{-\infty}^{0}\left(\frac{-1}{\Omega+\varsigma\omega}\right)\frac{\partial n_{F}(\omega)}{\partial\omega}d\omega\right]\nn
    = &\frac{ne^{2}(h/2\pi)}{2m^\star\Omega}-\left(\frac{ne^{2}(h/2\pi)\varsigma}{2m^\star\Omega^{2}}\right)\int_{-\infty}^{\infty}\frac{\beta\left|\omega\right|e^{\beta\omega}}{\left(1+e^{\beta\omega}\right)^{2}}d\omega \nn
    =& \frac{ne^{2}(h/2\pi)}{2m^\star\Omega}-\left(\frac{ne^{2}(h/2\pi)\varsigma Y}{2m^\star\Omega^{2}}\right)k_{B}T.
    \label{eq:sigma-app}
\end{align}
Here, $Y=\int_{-\infty}^{\infty}\frac{\left|x\right|e^{x}}{\left(1+e^{x}\right)^{2}}dx\approx1.39$ is a constant. 

Substituting for the constants $\Omega$ and $\varsigma$, the linear-in-$T$ part of the resistivity, denoted as $\rho_L(T)$, at sufficiently low temperatures can be proved to be inversely proportional to square of the Kondo coupling:
\begin{align}
    \rho_L(T)=\frac{m^\star}{ne^{2}}\left(\frac{Y}{\pi N j _{K}^{2}}\right)\frac{k_{B}T}{(h/2\pi)}\propto \frac{1}{j_K^2}.
    \label{eq:rho-linear}
\end{align}
In Eq. (\ref{eq:rho-linear}), we have replaced the density of states with the inverse of the conduction-electron bandwidth, namely $N_0 = \Lambda^{-1}$. 

At the QCP, we fix the dimensionless Kondo coupling $j_K$ in Eq. (\ref{eq:rho-linear}) at its fixed-point value, which is obtained via the RG analysis (see Supplementary Notes, section \ref{app:rg} and Ref. \cite{Chang-PRB-SM,Chang-SSC-PRB}):   $j_{K}^{2} \to \left( j_K^\star\right)^2=\eta \equiv d-z$ with $d$ beig the spatial dimension and $z$ being the dynamical exponent, which can be regarded as the effective dimension of time. This leads to 
\begin{align}
    \alpha \equiv \frac{Y}{\pi N\eta}.
\end{align}
Note that the value of $\eta$ is treated as a fitting parameter.

Taking the physical SU(2) limit (namely $N=1$), our result gives $\alpha \approx 0.98$ with $\eta \approx 0.45$ for the Nd-doped CeCoIn$_5$\cite{Chang-SSC-PRB}.

\section{Supplementary Notes: The electronic specific heat coefficient}
\label{supp:gamma}
The electronic specific heat coefficient in the strange metal state is mainly contributed from the quadratic Gaussian fluctuation of the $\hat{{\chi}}_\mbd{k}$ field, the Fourier component of the fluctuation of $\chi_{i}$ in Eq. (\ref{eq:HS}), whose dynamics is described by the Hamiltonian $H_G$, given by 
    \begin{align}
    	H_G = \sum_{\mbd{k}} \hat{{\chi}}_\mbd{k} \left[\varepsilon_\chi (\mbd{k}) + m_\chi \right]\hat{{\chi}}_\mbd{k} 
    	\label{eq:quadratic-H}
\end{align} 
with $\varepsilon_{\chi}(\mbd{k})$ being the dispersion of the $\hat{\chi}_\mbd{k}$ field which is obtained via calculating its leading-order self-energy correction and $m_\chi$ being the effective mass of $\hat{\chi}$. Note that, at the bare level, the $\hat{\chi}$ field does not acquire dynamics.

To proceed, we evaluate the ensemble average of internal energy $\bar{E}_G$ for $\hat{{\chi}}_\mbd{k}$, given by
  \begin{align} 
         \bar{E}_G =  \sum_\mbd{k} \frac{\varepsilon_\chi (\mbd{k})}{e^{\beta \left[\varepsilon_\chi (\mbd{k}) + m_\chi \right]}-1}
           =   \,W_\chi \, \int_{m_{\chi}}^{\Lambda}  \, d \varepsilon \, \frac{\varepsilon^{1+\eta/2}}{e^{\beta \varepsilon}-1} .
         \label{eq:InEnergy}
     \end{align}
     Here, $V$ denotes the system volume, $\beta \equiv 1/T$, $\Lambda$ is the upper energy cutoff scale and 
     \begin{align}
     	W_\chi \equiv  \frac{V\Omega_d}{2} \left(\frac{m^\star \lambda}{\pi N_0 J_K^2} \right)^{1+\eta/2}
     	\label{eq:W-chi}
     \end{align}
denotes the density of states for the $\hat{{\chi}}_\mbd{k}$ field with $\Omega_d \equiv 2 \pi^{d/2} / \Gamma(d/2)$ being the solid angle of a $d$-dimensional sphere and $\Gamma(n)$ being the gamma function with argument $n$. The specific heat coefficient can be obtained as
 \begin{align}
    \gamma(T)\equiv  \frac{1}{T}\frac{\partial \bar{E}_G}{\partial T} = \frac{W_\chi}{4T^3} \int^\Lambda_{m_\chi} \frac{\varepsilon^{2+\eta/2}}{\sinh^2\left( \frac{\varepsilon}{2T}\right)}d\varepsilon.
    \label{eq:gamma-unscaled}
 \end{align}

However, the parameters shown in Eqs. (\ref{eq:W-chi}) and (\ref{eq:gamma-unscaled}) such as the effective mass $m_\chi$, $\lambda$ or $J_K$ are at the bare level. Reasonable results of specific heat coefficient are obtained after all the bare parameters are replaced with the rescaled ones: $J_K\to J_K e^{-\eta l/2}$, $m_\chi \to m_\chi e^{2l}$ and $\lambda \to \lambda e^{\eta l/2}$. Note that, we also need to rescale temperature, given by $T(l) \equiv T_l = T e^{zl}$ with $l$ being the scaling parameter and $z$ being the dynamical exponent \cite{Hertz-RG,Millis-RG}. Since the conduction band does not get renormalized, thus $m^\star$ remains un-rescaled. Here, the density of states for the conduction electrons is expressed as $N_0 \sim m^\star k_F^{d-2}$, and is thus scale invariant at two dimension. Details of the rescaling of the parameters and temperature near the QCP are given in Ref. \cite{Chang-PRB-SM}. 

The specific heat coefficient contributed from the fluctuations of the Kondo hybridization sector should be evaluated near the critical fixed point at which $m_\chi$ is renormalized to of order of one, $m_\chi(l=l_0) \sim O(1)$ where $l_0$ denotes the scaling factor of renormalization at which the above relation is satisfied, i.e. $m_{\chi }(l_0) \sim O(1) \Rightarrow 1 = m_\chi e^{z \, l_0} = m_\chi  \xi^z$ with $m_{\chi} $ here being the bare mass of the $\hat{{\chi}}$-field, $\xi$ being the correlation length, and $T (l=l_0) = T e^{z l_0} \sim T \xi^z \sim T/T_{LFL}$ with $T_{LFL}$ being the Fermi-liquid crossover. Taking the renormalized parameters into accounting, the (renormalized) specific heat coefficient $\gamma(T) \equiv C_V /T$ is obtained by the definition 
 \begin{align}
     \frac{C_V}{T} 
       & =  \, e^{-\eta  l_0} \, W_\chi \, \int^{\Lambda}_{m_\chi^l} \,
d \varepsilon_l \,\,   \frac{\beta_l^{2}\, \varepsilon_l^{2+\eta/2} \,e^{\beta_l \varepsilon_l}}{\left( e^{\beta_l \varepsilon_l}-1 \right)^{2}} \Bigg|_{l = l_0} \nn
      & = \, -\frac{1}{4} \, A(l_0) \,\left(\frac{T}{T_{LFL}}\right)^{\eta/2}\,\int^{m_\chi \Lambda /T}_{m_\chi /T} \, dx \, \frac{x^{2+\eta/2}}{\sinh^{2} (x/2)}
      \label{eq:gamma}
   \end{align}
with $x \equiv \varepsilon_l /T_l$, $m_\chi^l \equiv  m_\chi(l)$, $\varepsilon_l \equiv \varepsilon(l)$. In Eq. (\ref{eq:gamma}), $A(l_0) = e^{-\eta l_0} W_\chi$ is a non-universal parameter while the integral part is a universal scaling function. $A(l_0)$ and $m_\chi$ in Eq. (\ref{eq:gamma}) are treated as numerical fitting parameters. In the above derivations, the relations $e^{l_0} = \xi$ and $\xi^z \sim T_{LFL}^{-1} \sim (g-g_c)^{-1} \sim (J_K - J_K^{\ast})^{-1}$ have been used since the RG flows along the direction of $J_K$ with $J_H$ fixed at $J_H^{\ast}$ dominates the critical properties.

We now examine the scaling behavior of the prefactor $A(l_0)$, which will be shown to exhibit a power-law behavior of the specific heat coefficient in $g-g_c $ near the quantum critical point. Replacing the bare $\lambda$ and $J_K$ in Eq. (\ref{eq:W-chi}) with the resclaed ones shown above and using the relation $e^{l_0} \sim (g-g_c)^{-1/2}$ yields
\begin{align}
    A(l_0)= &  e^{l_0} W_\chi (l=l_0)\propto e^{l_0} \left( \frac{\lambda }{J_K^2 }\Bigg|_{l=l_0}\right)^{1+\eta/2} \nn
    & \propto (g-g_c)^{\eta/2} \left[ \frac{(g-g_c)^{-\eta/4}}{(g-g_c)^{\eta/2}}\right]^{1+\eta/2}\nn 
    & = (g-g_c)^{-\beta}
\end{align}
with $\beta = 3\eta^2/8 + \eta/4$. This shows that the specific heat coefficient exhibits a power-law behavior in $g-g_c$ with fixed temperature near the QCP.


\end{document}